\documentclass[12pt,preprint]{aastex}
\usepackage[yyyymmdd,hhmmss]{datetime}
\usepackage{amsmath}
\usepackage{gensymb}
\usepackage{natbib}

\shorttitle{WRs in the Magellanic Clouds}

\shortauthors{Massey et al.}

\begin{document}

\title{A Modern Search for Wolf-Rayet Stars in the Magellanic Clouds. III. A Third Year of Discoveries\altaffilmark{*}}

\author{Philip Massey\altaffilmark{1,2}, Kathryn F. Neugent\altaffilmark{1,2}, and Nidia Morrell\altaffilmark{3}}

\altaffiltext{*}{This paper includes data gathered with the 1 m Swope and 6.5 m Magellan Telescopes located at Las Campanas Observatory, Chile.}
\altaffiltext{1}{Lowell Observatory, 1400 W Mars Hill Road, Flagstaff, AZ 86001; phil.massey@lowell.edu; \\kneugent@lowell.edu.}
\altaffiltext{2}{Also Department of Physics and Astronomy, Northern Arizona University, Box 6010, Flagstaff, AZ 86011-6010.}
\altaffiltext{3}{Las Campanas Observatory, Carnegie Observatories, Casilla 601, La Serena, Chile; nmorrell@lco.cl.}

\begin{abstract}

For the past three years we have been conducting a survey for WR stars in the Large and Small Magellanic Clouds (LMC, SMC).  Our previous work has resulted in the discovery of a new type of WR star in the LMC, 
which we are calling WN3/O3.  These
stars have the emission-line properties of a WN3 star (strong N\, {\sc v} but no N\, {\sc iv}), plus the absorption-line properties of an O3 star (Balmer hydrogen plus Pickering He\, {\sc ii}, but no He\, {\sc i}).  Yet these stars are 15$\times$ fainter than an O3~V star would be by itself, ruling out these being WN3+O3 binaries. Here we report the discovery of two more members of this class, bringing the total number of these objects to 10, 6.5\% of the LMC's total WR population.  The optical spectra of nine of these WN3/O3s are virtually indistinguishable from each other, but one of the newly found stars is significantly different, showing a lower excitation emission and absorption spectrum (WN4/O4-ish). In addition, we have newly classified three unusual Of-type stars, including one with a strong C\,{\sc iii} $\lambda 4650$ line,  and two rapidly rotating ``Oef" stars.  We also ``rediscovered" a low mass x-ray binary, RX J0513.9-6951, and demonstrate its spectral variability.  Finally, we discuss the spectra of ten low priority WR candidates that turned out not to have He\, {\sc ii} emission. These include both a Be star and a B[e] star.

\end{abstract}

\keywords{galaxies: stellar content --- galaxies: individual (LMC, SMC) --- Local Group --- stars: evolution --- stars: Wolf-Rayet}

\section{Introduction}
\label{Sec-intro}

For many years, our knowledge of the Wolf-Rayet (WR) population of the Magellanic Clouds (MCs) was considered essentially complete: twelve WRs were known in the SMC \citep{MasseyWRSMC} and 134 were known in the LMC  \citep{BAT99}.  These stars had been found by a combination of general objective prism surveys, directed searches, and accidental discoveries by spectroscopy \citep[see][]{MasseyRev13, MasseyPots}.   However, over the years several additional WRs were found in the LMC, culminating in our own discovery of a very strong-lined WO-type \citep{NeugentWO}, only the second known example of this rare type of WR in the LMC.    This discovery prompted us to begin a multi-year survey of both the SMC and LMC in an effort to obtain a complete census of their Wolf-Rayet population.  In part this was motivated in terms of finding a more accurate value for the relative number of WC- and WN-type WRs, as this provides a key test of the evolutionary models (see, e.g., \citealt{MJ98}, \citealt{MeynetMaeder05}, \citealt{NeugentM31}).  And, such a survey was timely, given our improved knowledge of the populations of other evolved massive stars in the Magellanic Clouds, such as yellow and red supergiants \citep{NeugentSMC, NeugentLMC}, and on-going improvements in massive star models \citep{Sylvia,JJ}.

It was entirely possible, of course, that our survey would fail to find much of interest.  Instead, in the first year we discovered nine more WRs in the LMC \citep[][hereafter Paper I]{MasseyMCWRI}.  More interesting than the numbers, however, were the type of WRs we found:  six had spectra that were unlike those of any previously observed.  They were also somewhat fainter in absolute visual magnitude ($M_V=-2.5$ to $-3.0$) than the previously known WRs.  Their emission-line spectra were dominated by N\,{\sc v} $\lambda\lambda$4603,19, $\lambda$4945 and He\,{\sc ii} $\lambda$4686, and no trace of N\,{\sc iv} $\lambda$4058, implying a WR class of WN3. However, they also showed He\,{\sc ii} and Balmer absorption spectra with no trace of He\,{\sc i}, characteristic of an O3~V star.   Yet, these stars were 10$\times$ too faint to be WN3+O3~V binaries, as a typical O3~V star has $M_V\sim -5.5$ \citep{Conti88,WalbornO2}.   Our preliminary modeling demonstrated that we could reproduce both the emission and absorption lines with a single set of physical parameters \citep[Paper I; ][]{NeugentWN3O3}, lending further credence to these being single objects.  (Our limited number of repeat observations also failed to detect any radial velocity variations.) In our second year of the survey we detected two more of these stars \citep[][hereafter Paper II]{MasseyMCWRII}, which we are calling WN3/O3s.    In addition, our survey has found the third known WO-type star in the LMC and four other WNs (including two WN+O binaries, an O3.5If*/WN5 star, and a WN11), two rare Of?p stars (magnetically braked oblique rotators; see \citealt{WalbornOf?p})\footnote{See the recent followup study of these two stars plus other members of this class by \cite{Yael}}, an Onfp star \citep[see][]{WalbornOnfp}, a possible B[e]+WN binary, four Of-type supergiants, and a peculiar emission-line star (the nature of which eludes us), and other, more normal, early-type stars (Papers I and II).  

Here we report on the third year of discoveries, which include two additional members of the WN3/O3 class in the LMC, bringing the number of this class to 10, 6.5\% of the LMC's WR population, which now stands at a total of 154 stars. One of these WN3/O3 stars is not like the others, in that it shows somewhat lower excitation emission and absorption lines. We report the discovery of three Of stars (all unusual),   and revisit a low mass x-ray source previously known to have a very broad He\,{\sc ii} $\lambda$4686 component.  We also classify 10 low ranked candidates that failed to have He\,{\sc ii} $\lambda$4686, including one Be and one B[e] star.

\section{Observations and Reductions}

The imaging part of our survey is being conducted using the Swope 1-m telescope on Las Campanas.  Complete details are given in Papers I and II; here we summarize our observing and
reduction procedures. The
current e2v camera provides a 29\farcm8(EW) $\times$ 29\farcm7(NS) field of view with 4110$\times$4096 15$\mu$m pixels, each subtending 0\farcs435. Our exposure times are
300 s through each of three interference filters, a {\it WC} filter centered on C\,{\sc iii} $\lambda$4650, the strongest optical line in WC- and WO-type WRs, a {\it WN} filter centered on
He\,{\sc ii} $\lambda$4686, the strongest optical line in WN-type WRs, and a continuum filter ({\it CT}), centered at 4750\AA.  All three filters have a 50\AA\ bandpass (full width at half maximum, FWHM).   We were initially assigned ten nights on the Swope (UT) 2015 Nov 15-24.  However, since some of that time was lost due to clouds, Carnegie was kind enough to assign us five additional nights, Dec 25-29, which had been unscheduled.    Typical image quality on both runs ranged from 1\farcs3-1\farcs9.  There were 93 fields observed (or reobserved) in the LMC, and 21 in the SMC in total during this third year of our survey.  There is about a 1\arcmin\ overlap between adjacent fields, and so the total new coverage is  19.9 deg$^2$ in the LMC, and 4.5 deg$^2$ in the SMC.

For calibration purposes, ten bias frames were taken daily.  When conditions were clear, 3-5 sky flats were obtained in bright twilight through each filter, dithering the telescope between exposures in order to allow us to filter out any stars when we combined frames. Since these twilight exposures were short (a few seconds), it was necessary to correct each of the exposures for the iris pattern of the camera shutter.  The details of this procedure are given in Paper II.   Since the CCD is read out through four amplifiers, each quadrant is treated separately in the preliminary reductions.  Overscan and bias structure (which is practically non-existent) were removed.  After these additive corrections were made, the counts in each quadrant were corrected for slight non-linearities in the CCD amplifiers as discussed in Paper II.  Each quadrant was flat-fielded, and then the four sections were recombined.  Accurate (0\farcs5) coordinates are obtained through the use of the ``astrometry.net" software \citep{Lang}.  

The frames were then analyzed by first running aperture photometry on each image to identify stars that were significantly brighter in one of the two emission-line filters than in the continuum image.  In addition, image subtraction was performed with the High Order Transform of Point-spread function ANd Template Subtraction ({\sc hotpants}) software described by \cite{Becker}.  The resultant {\it WC-CT} and  {\it WN-CT} images were examined by eye to identify WR candidates.    The combination of the two techniques has proven very effective, as shown in Papers I and II.  

This procedure was carried out under some time pressure as we had a single night (UT 11 Jan 2016) allocated on the Baade 6.5-m Magellan telescope for follow-up spectroscopy.  Despite the short time between the December imaging and January spectroscopy time, we had successfully reduced and analyzed most of the fields, with 15 WR candidates at various significance levels. 

The spectroscopic observations took place with the Magellan Echellette Spectrograph (MagE) mounted on a folded port of the Baade.  The instrument is described in detail by \cite{MagE}. 
MagE provides complete coverage from the atmospheric cutoff ($\sim$3200\AA) to $\sim$1$\mu$m.  We used it with a 1\arcsec\ slit, which then yielded a resolving power $R$ of 4100. 
The night was clear, and the seeing was 0\farcs6-0\farcs7.   As we described in Paper II (and in more detail in \citealt{MasseyHanson}),
the pixel-to-pixel sensitivity variations are quite small, and we have found we can achieve better results by not flat-fielding the data in the blue.  The data from this night were flat-fielded in the red, primarily to remove fringing. After extraction, wavelength calibration, and fluxing, the individual orders were combined.  Our typical signal-to-noise (S/N)  in the blue classification region is 100-150 per spectral resolution element.

Our initial spectrum of LMCe058-1 was intriguing but quite noisy, and we were fortunate to be able to squeeze in additional
MagE observations of the source on three additional nights throughout the following months, as described further below.  
Analysis of a few remaining fields not fully analyzed before our January night  revealed one other high-significance WR candidate (LMCe113-1), which was observed with MagE on UT 29 March 2016 at the start of the night.

\section{Discoveries}

Our spectroscopy identified six stars among our candidates that have He\,{\sc ii} $\lambda$4686 emission.  Ten other candidates proved to be
(mainly) B-type stars, not of immediate interest to us, but discussed further below.  This success rate is similar to our second year (Paper II).  We discuss our ``winners" and ``losers" below.

\subsection{Newly Found Wolf-Rayet Stars}

Our most exciting discovery was identifying two new members of the class of WRs that 
we are calling WN3/O3s (Papers I and II).  Other members of this class
show the strong N\,{\sc v} $\lambda\lambda$4603,19 doublet, N\, {\sc v} $\lambda$4945 and He\,{\sc ii} $\lambda$4686 emission lines characteristic of a WN3 star, and a He\,{\sc ii} and Balmer line absorption spectrum typical of an O3 star.  A composite spectrum can be immediately ruled out as these stars have $M_V=-2.3$ to $-3.0$, 15$\times$ fainter than an O3~V would be by itself ($M_V\sim -5.5$, \citealt {Conti88}) and even slightly fainter than a normal WN3
star would be by itself ($M_V\sim -3.8$, \citealt{PotsLMC}).

\begin{figure}
\epsscale{1.0}
\plotone{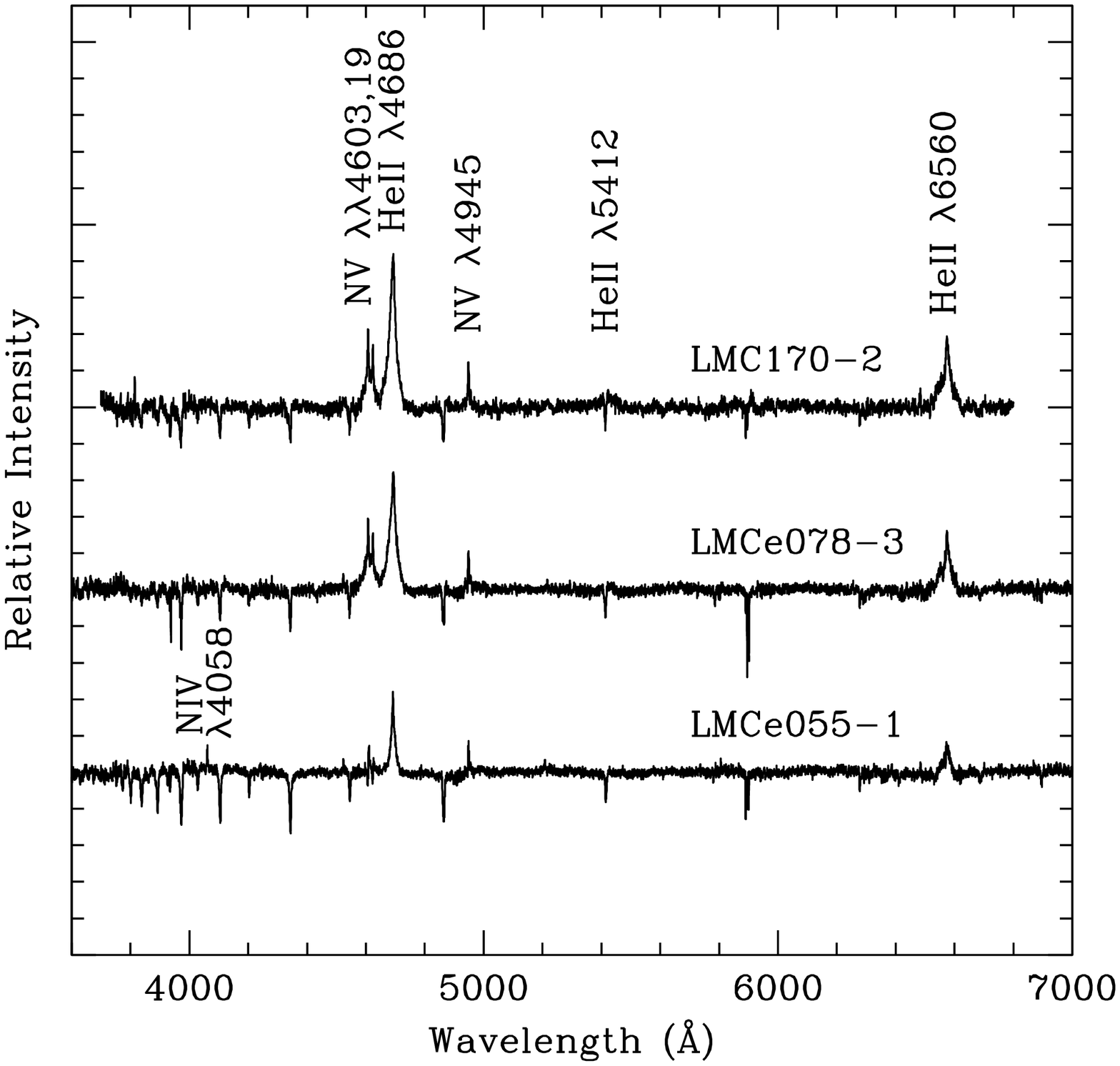}
\caption{\label{fig:O3s} Comparison of the spectra of our two newly found WN3/O3 stars with the prototype of this class, LMC170-2 (Paper I). 
LMe078-3 is very similar, but LMCe055-1 shows lower emission excitation (N\,{\sc iv} $\lambda$4058, which is missing in the other two) and the N\,{\sc v} $\lambda\lambda$4603,19 has a P Cygni profile.  Comparison of the absorption lines is better seen in Fig.~\ref{fig:O3sblue}.
}
\end{figure}

\begin{figure}
\epsscale{0.48}
\plotone{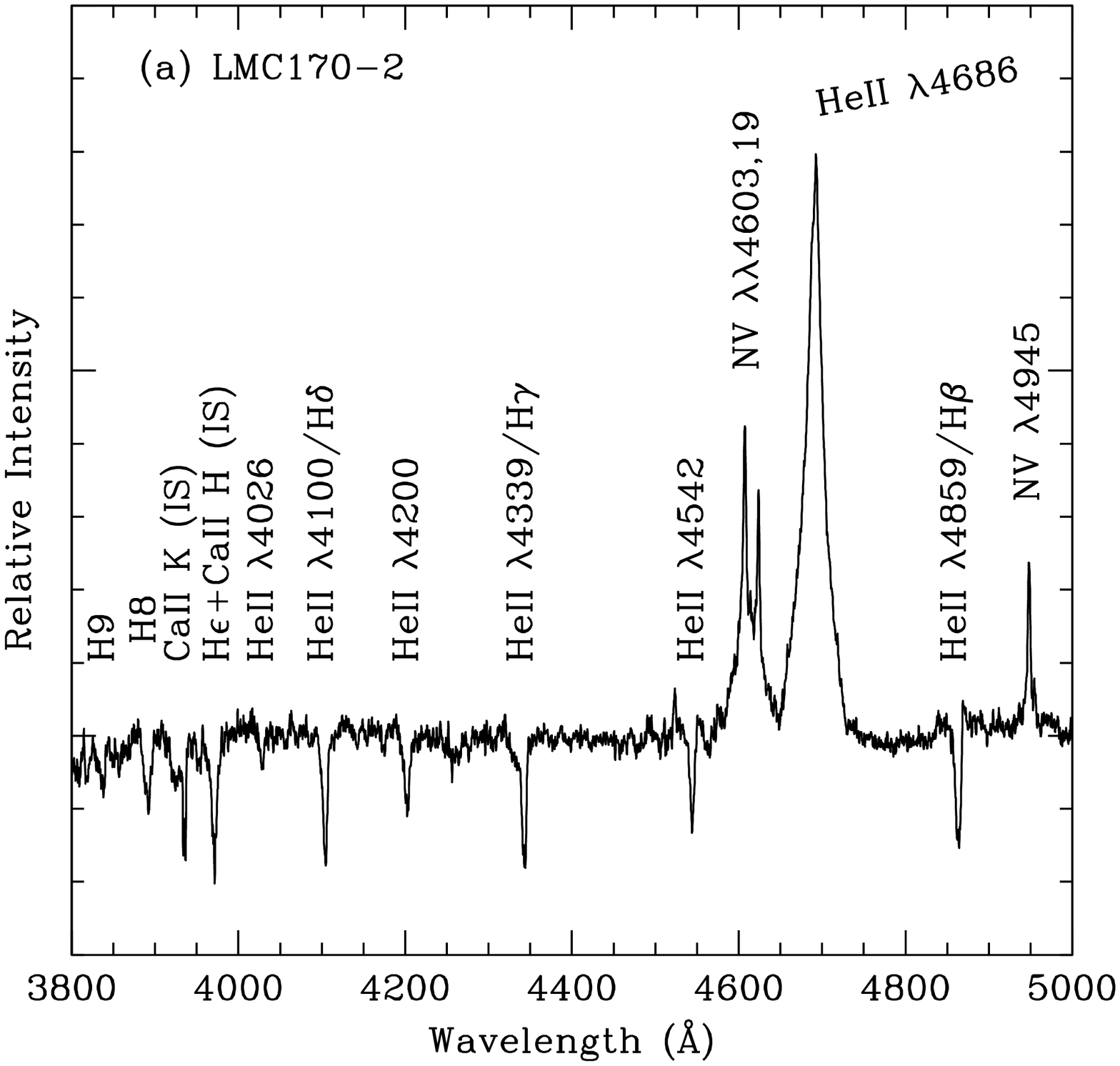}
\plotone{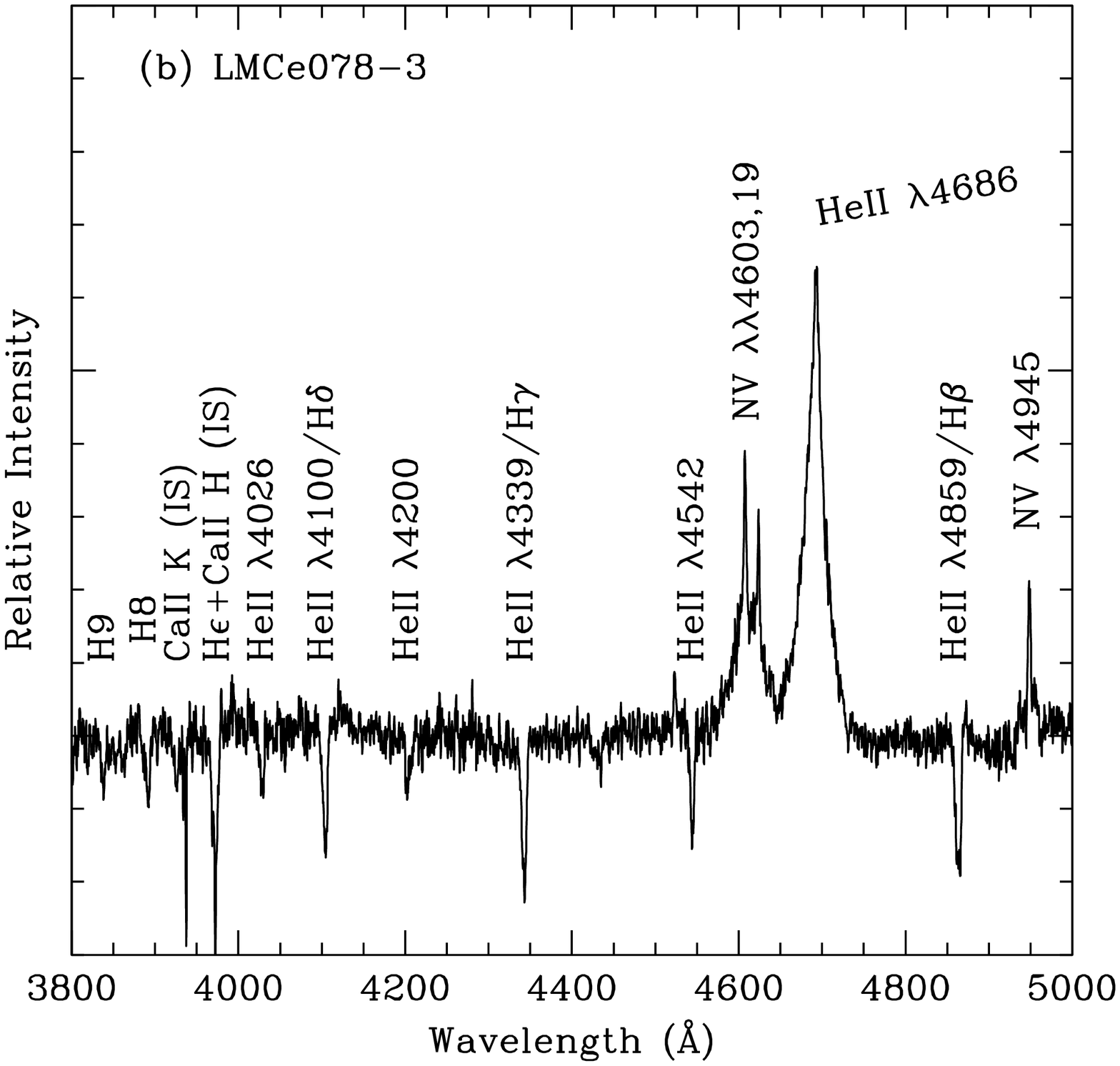}
\plotone{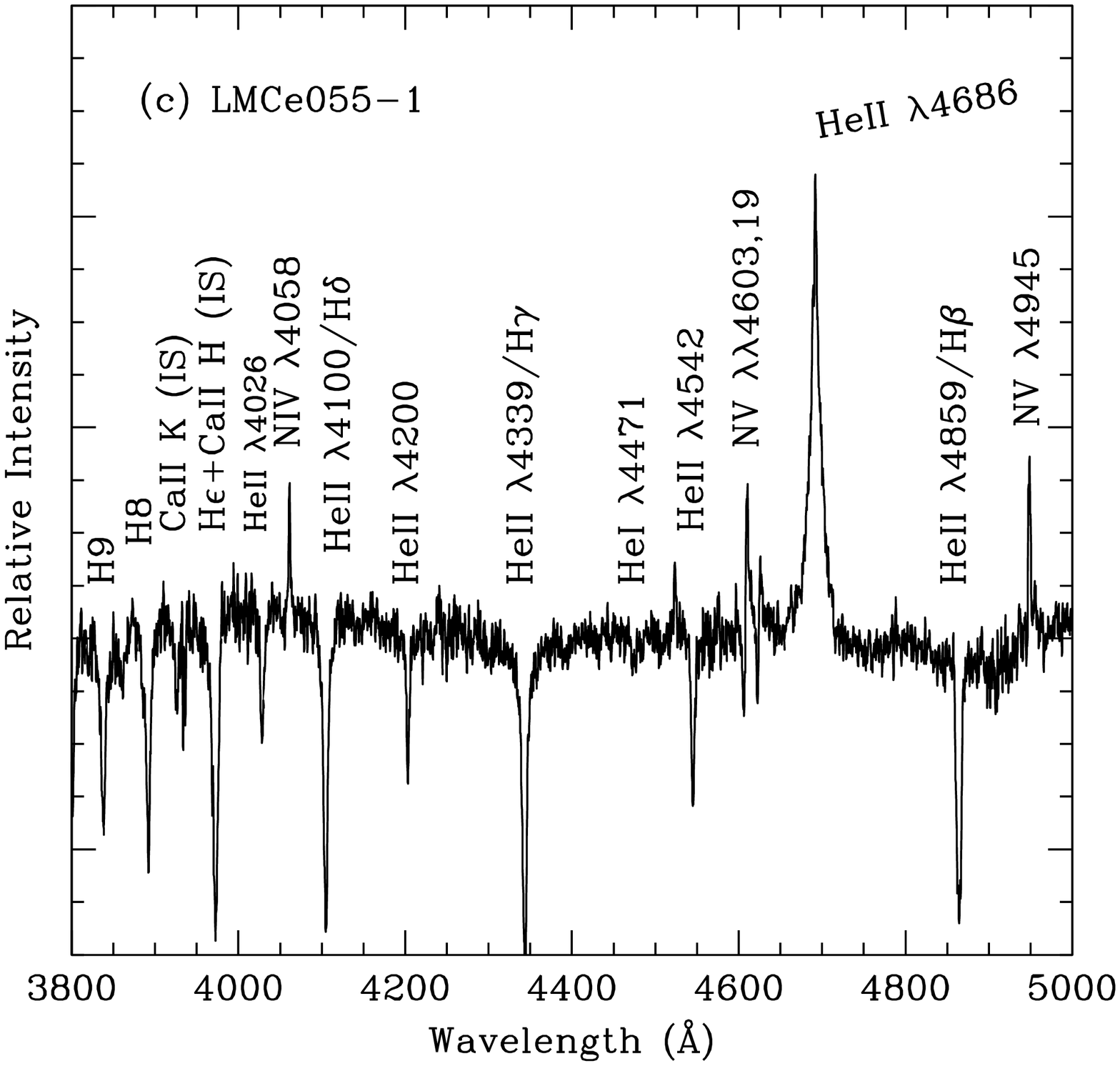}
\caption{\label{fig:O3sblue} Expansion of the blue region of the spectra shown in Fig.~\ref{fig:O3s}. Here we see how similar the absorption spectrum is in LMCe078-3 compared to that of LMC170-2, while LMCe055-1 show weak He\,{\sc i} $\lambda 4471$ absorption.}
\end{figure}

We list the properties of these two WN3/O3 stars in Table~\ref{tab:WRs} and illustrate their spectra in Figs.~\ref{fig:O3s} and \ref{fig:O3sblue}. One thing we have been strongly struck by is how similar all the previous members of this group have been; their spectra are nearly indistinguishable; see, e.g., Fig.~7 in Paper I and Fig.~3 in Paper II.
LMCe078-3 is now the ninth example, with a spectrum nearly identical to one of the prototypes of this class, LMC170-2, also illustrated here in
Figs.~\ref{fig:O3s} and \ref{fig:O3sblue}.

However, the spectrum of our newly discovered tenth member of the WN3/O3 class, LMCe055-1, is substantially different.  Both the emission and the absorption are indicative of a lower excitation temperature.  The spectrum shows N\,{\sc iv} $\lambda$4058 emission, which is completely missing from the other
members of this group.  Furthermore, weak He\,{\sc i} $\lambda$4471 is visible in absorption.  The N\, {\sc v} $\lambda\lambda$4603,19 doublet lines have P Cygni profiles, with a strong absorption component to the blue side of the emission.  We would more properly call this star a
WN4/O4 rather than an WN3/O3!  

\begin{figure}
\epsscale{1.0}
\plotone{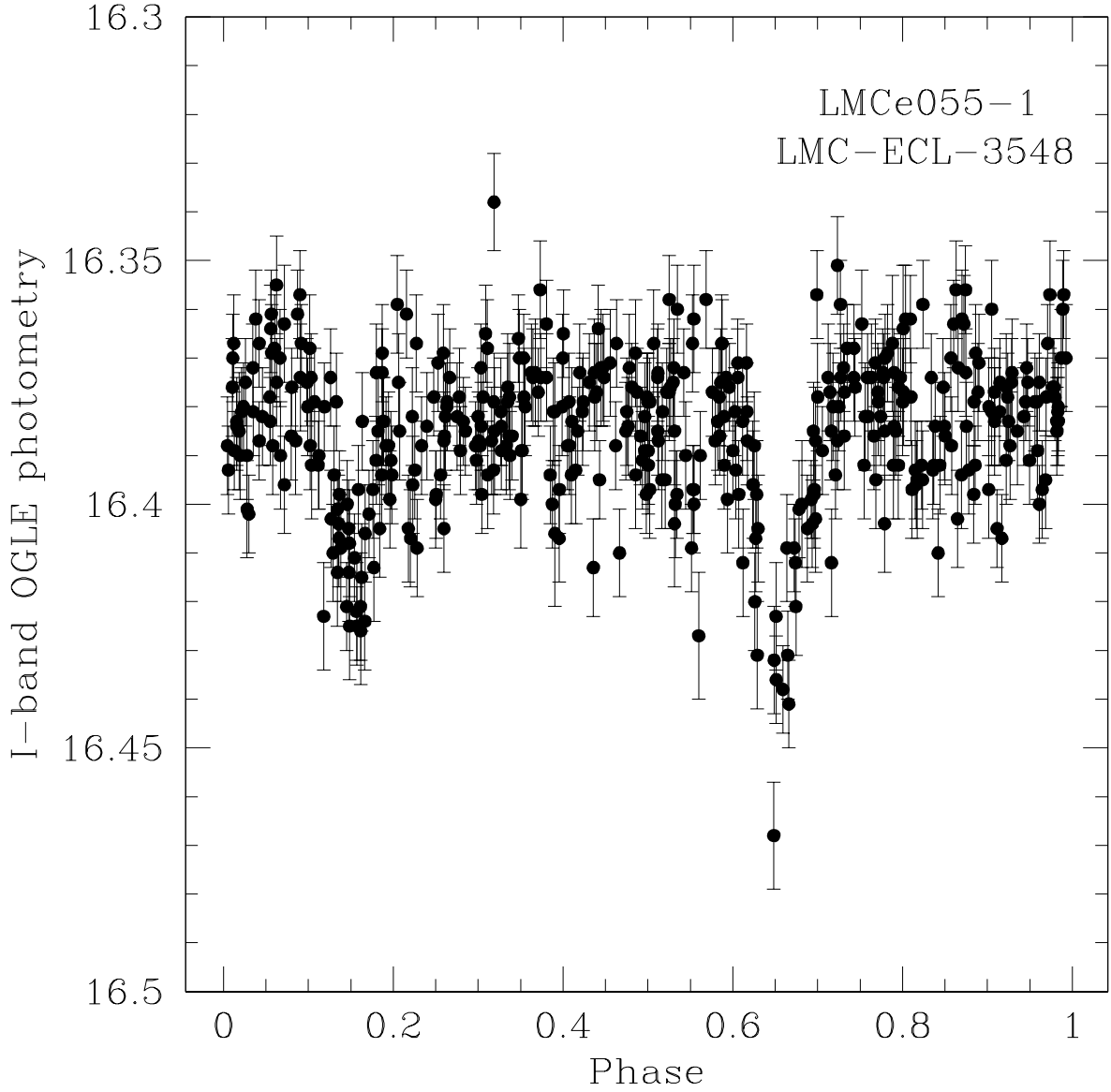}
\caption{\label{fig:OGLE} Light-curve of LMCe-055-1 (OLGE LMC-ECL-3548). The I-band photometry is from OGLE \citep{OGLE}, and has been phased with their period of
2.159074 days.}
\end{figure}

Further complicating the picture is the fact that this star was identified as an eclipsing binary by OGLE with a 2.159074 day period \citep{OGLE}.
Might this star simply be a normal WN4+O4~V binary?  We can rule this out immediately using the same argument as for ``normal" WN3/O3 stars: as shown in Table~\ref{tab:WRs} has an absolute
visual magnitude of only $M_V=-2.8$, while an O4~V star is expected to have an $M_V\sim -5.5$. \citep{Conti88}.  Furthermore, the OGLE light
curve itself is inconsistent with the presence of a massive companion.   As shown in Fig.~\ref{fig:OGLE} there is no sign of the ellipsoidal variations
one sees in massive binaries with similarly short periods, such as DH~Cep \citep{DHCep}\footnote{We are indebted to our colleague Dr.\ Laura Penny for commenting on the OGLE light curve, and pointing out the implications of the comparison with DH~Cep.}.  We are continuing to investigate this star, including a radial velocity study.

\subsection{Other Emission Line Stars}

\subsubsection{Of-type stars}
Our survey is sufficiently sensitive that we detect many Of-type stars. Although such stars are significantly brighter than our WN3/O3s, the equivalent widths (EWs)  of their
He\,{\sc ii} $\lambda$4686 emission are sufficiently small that these are among the hardest stars for us to detect (see, e.g., Fig.~9 in Paper~II).  Here we discuss the discovery of three of these stars.

\begin{figure}
\epsscale{0.5}
\plotone{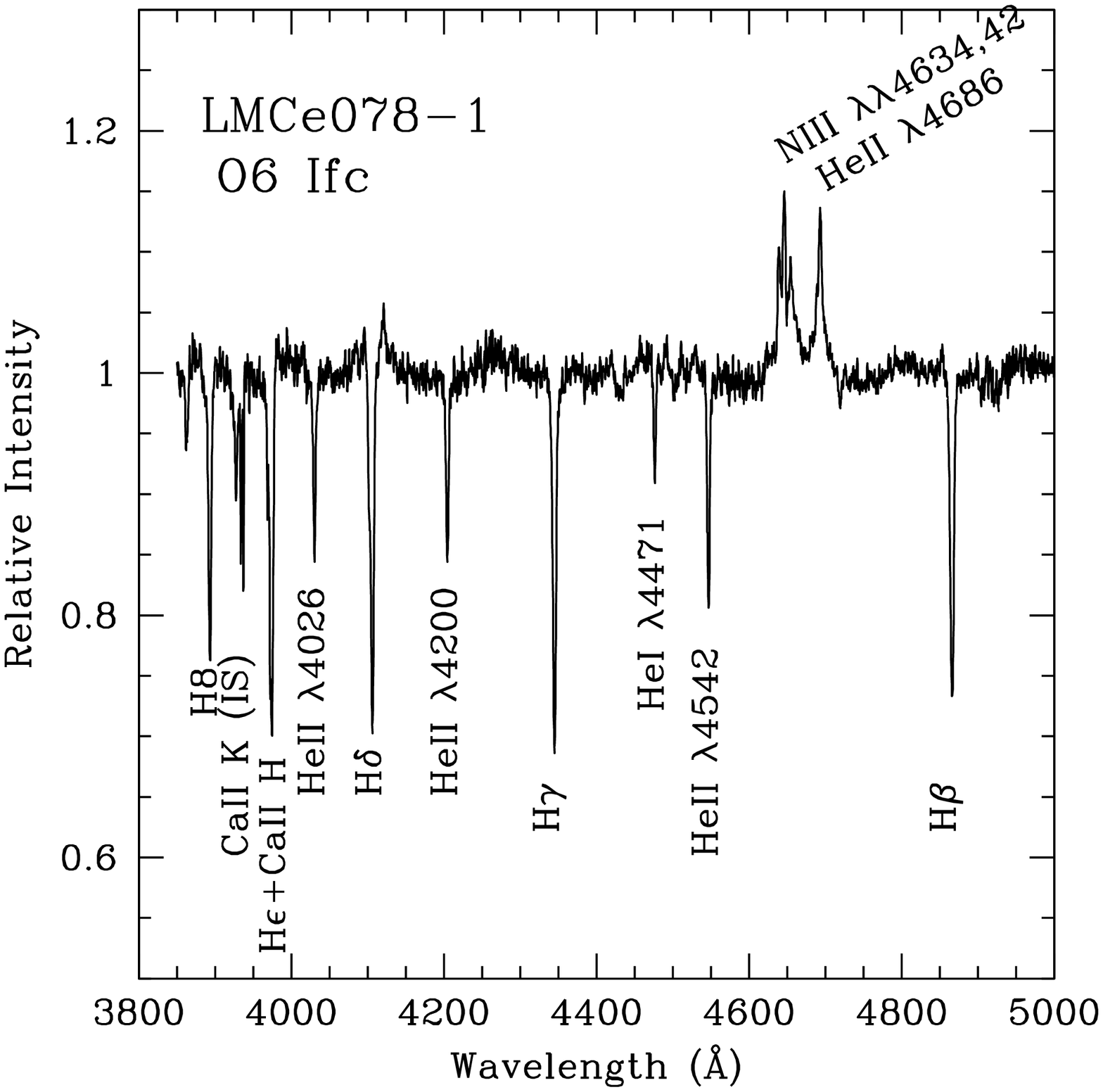}
\plotone{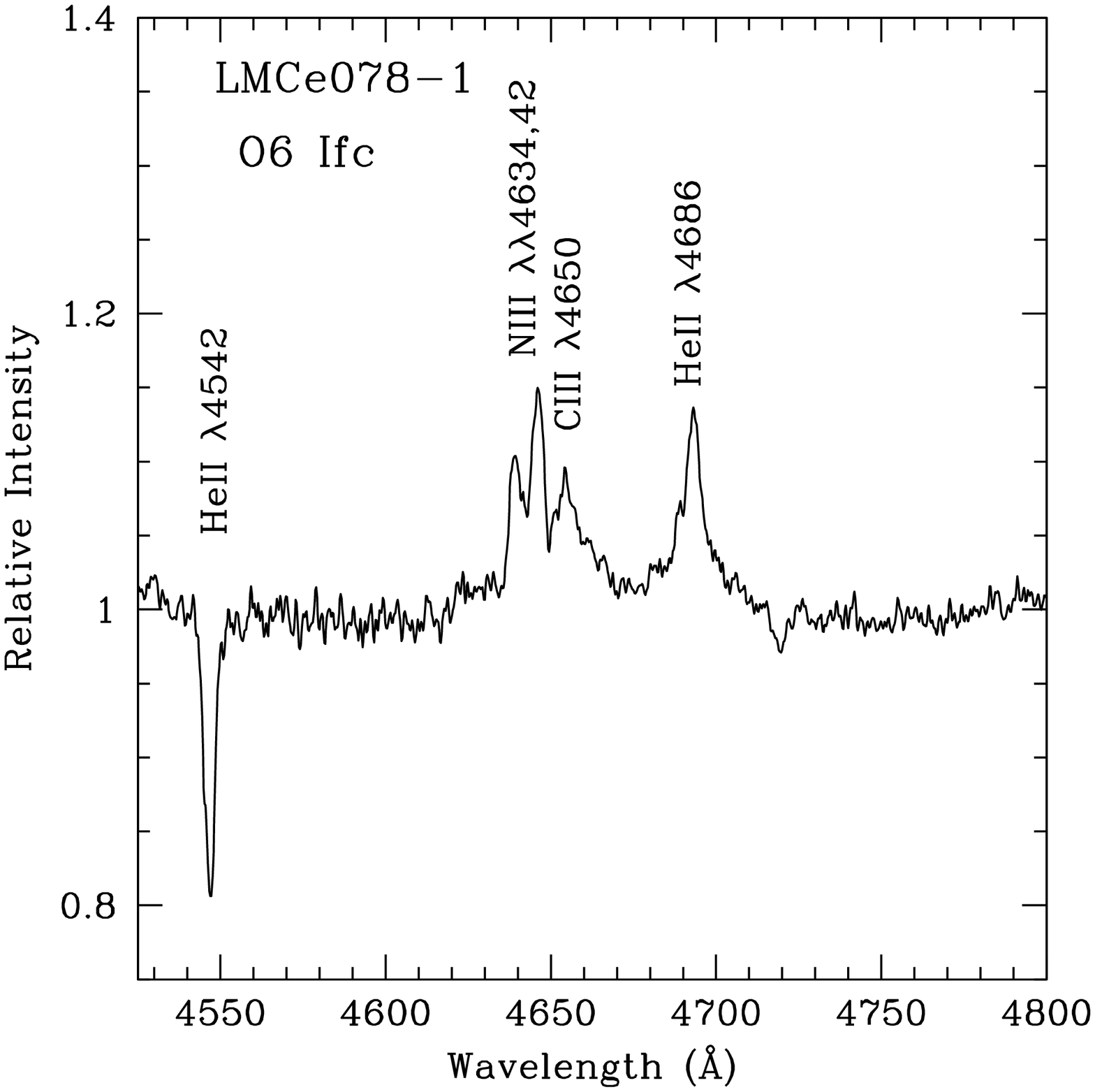}
\caption{\label{fig:LMCe0781} Spectrum of the O6 Ifc star LMCe078-1.  The presence of C\,{\sc iii} $\lambda$4650 necessitates the ``c" in
the luminosity designation.} 
\end{figure}

The spectrum of LMCe078-1 is shown in Fig.~\ref{fig:LMCe0781}. We classify the star as O6 Ifc, where the ``c" is required given the strong presence of C\,{\sc iii} $\lambda$4650 emission \citep{WalbornOf?p} in addition
to the strong N\,{\sc iii} $\lambda\lambda$4634,42 and He\,{\sc ii} $\lambda$4686 lines that result in the luminosity ``If" classification (.
Subsequent to our spectroscopy, but prior to this publication, \cite{Evans} also classified this star (their star \#271)
 as an O5.5~Iaf, in substantial
agreement with our classification here, although the presence of the C\,{\sc iii} emission went unnoticed. (The line is only marginally visible in the on-line version of their spectrum, due to their lower S/N.)

\begin{figure}
\epsscale{0.5}
\plotone{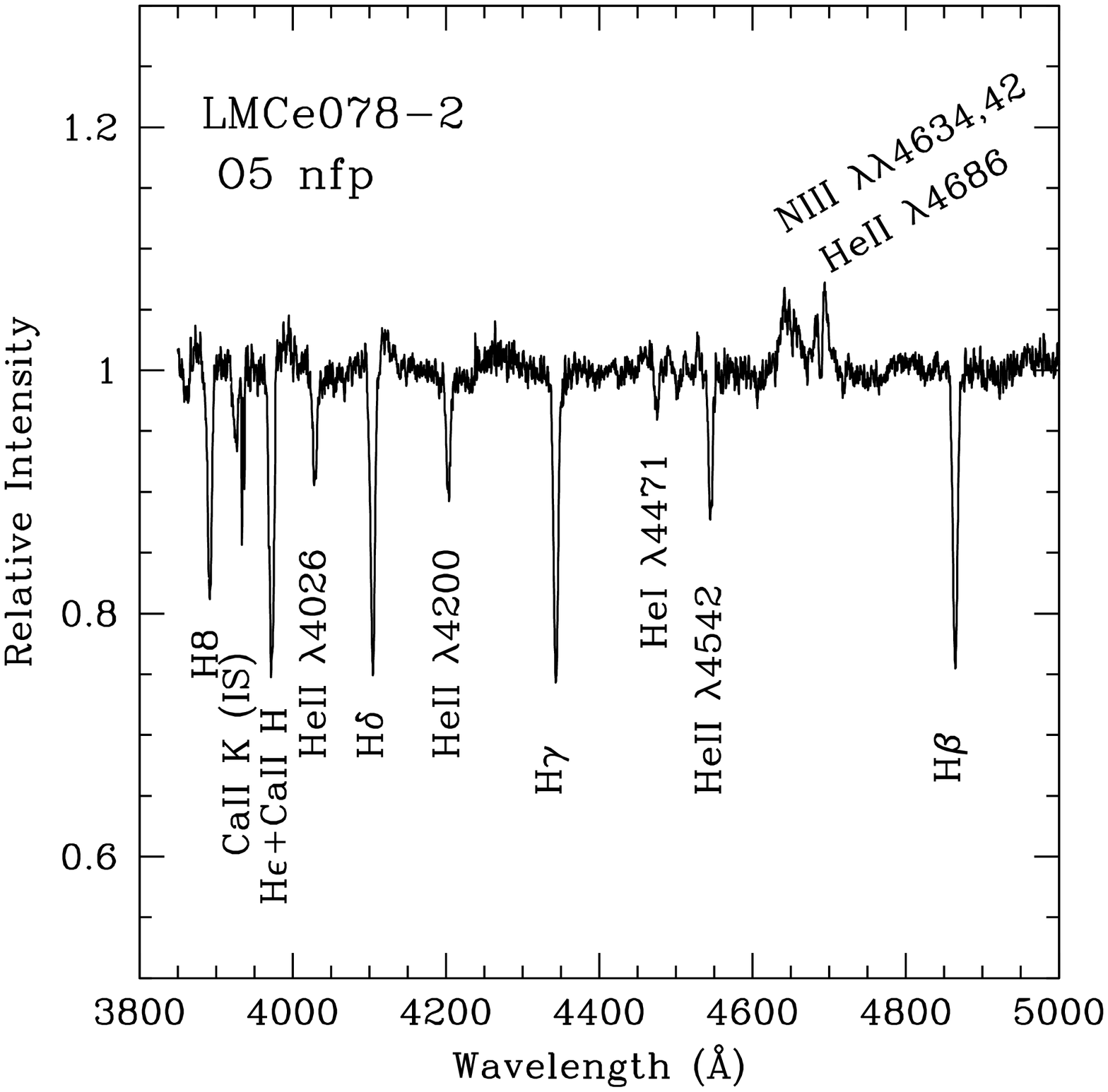}
\plotone{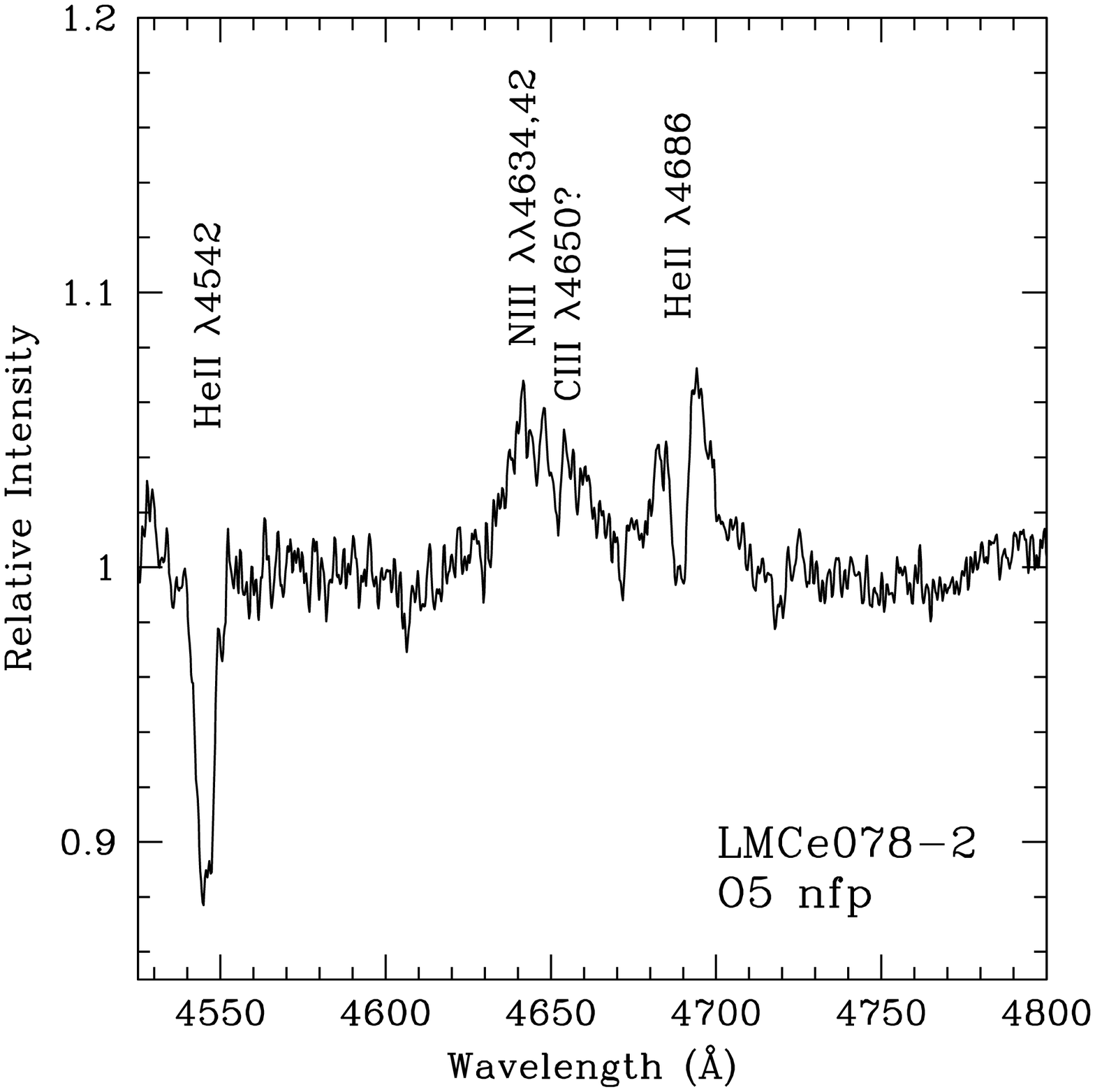}
\caption{\label{fig:LMCe0782} Spectrum of the O5 nfp star LMCe078-2.  Note the reversal (absorption) in the He\,{\sc ii} $\lambda$4686 line,
characteristic of the ``nfp" designation.}
\end{figure}

Another Of-type star, LMCe-078-2, is shown in Fig.~\ref{fig:LMCe0782}.  It too is unusual: He\,{\sc ii} $\lambda$4686 emission shows a central reversal (absorption),
and the absorption lines are extremely broad, with $v \sin{i} \sim$ 385 km~s$^{-1}$.  These are the classic characteristics of ``Oef" stars \citep{ContiOef},
also known as ``Onfp" stars \citep{Walborn73,WalbornOnfp,Sota}.  Based upon the relative He\,{\sc ii} and He\,{\sc i} absorption line strengths, we classify
the star as O5nfp.  Given the weakness of the emission, it is rather remarkable that we detected this star in the first place. 
Fortunately the N\,{\sc iii} $\lambda\lambda$4634,42 emission doublet is nicely within the bandpass of the {\it WC} filter, and so this star was considered of high significance as it was brighter in both emission-line filters relative to the continuum. 
In addition,  Onfp stars are known to have variable He\,{\sc ii} strengths, and possibly we detected this star when the line was stronger. We note that this star too may show signs of C\,{\sc iii} $\lambda$4650 as well, but the broadness of the lines precludes a definitive assessment. 

\begin{figure}
\epsscale{0.5}
\plotone{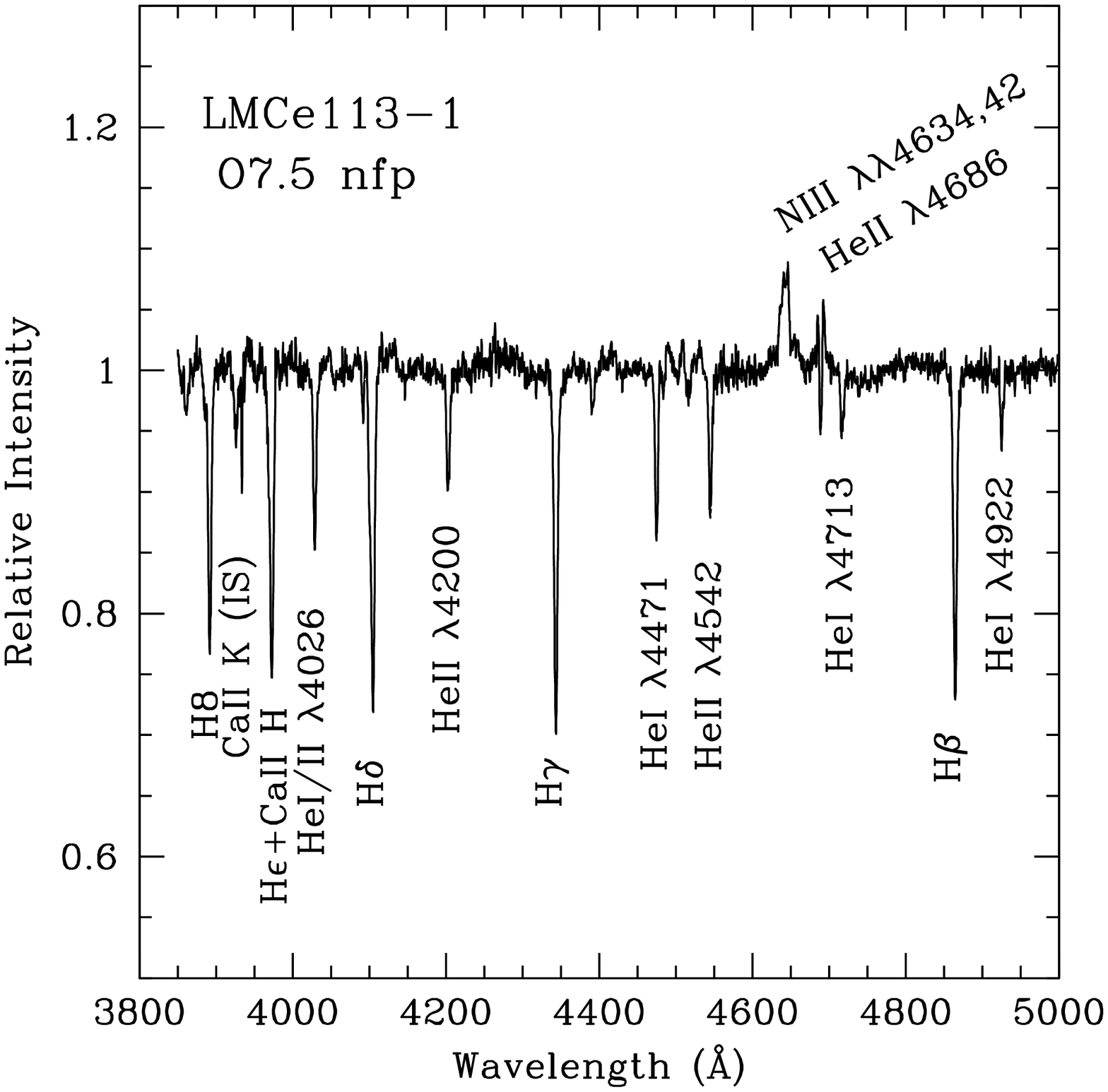}
\plotone{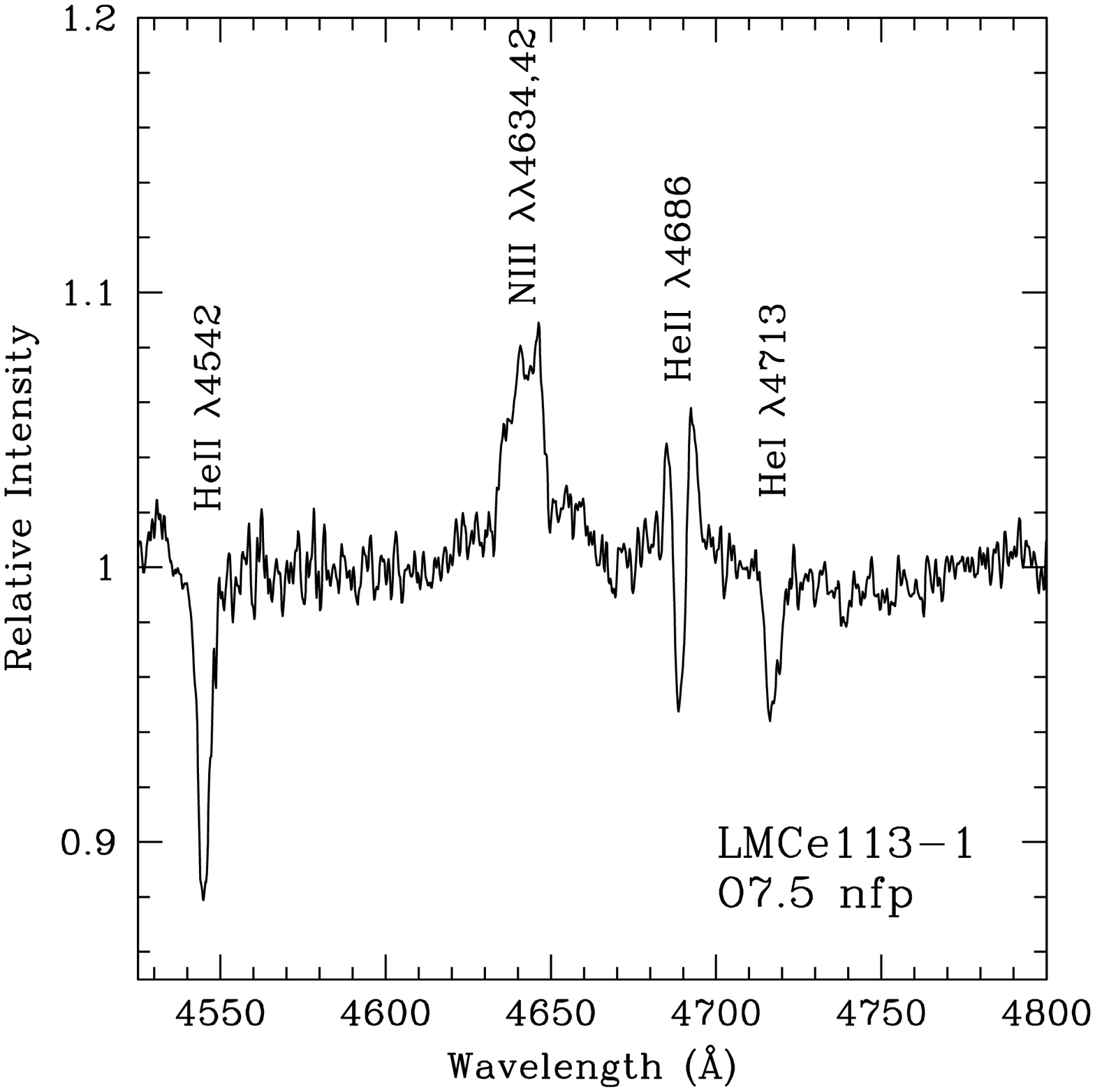}
\caption{\label{fig:LMCe1131} Spectrum of the O7.5 nfp star LMCe113-1.  Note the reversal (absorption) in the He\,{\sc ii} $\lambda$4686 line,
characteristic of the ``nfp" designation.}
\end{figure}

Our third Of-type star, LMCe113-1, is yet another Onfp star, which we classify as O7.5nfp.  As shown in Fig.~\ref{fig:LMCe1131}, He\,{\sc ii} $\lambda 4686$ has a central reversal.  The emission here is quite weak, we are again
surprised (but pleased) we detected it; without the N\,{\sc iii} $\lambda\lambda$4634,42 emission we probably would not have.    The absorption lines are quite broad, with
$v \sin{i} \sim$ 300 km~s$^{-1}$.  The star is identified in the \cite{Sanduleak} catalog as Sk$-67^\circ$3 according to Brian Skiff's on-line spectral catalog\footnote{http://cdsbib.u-strasbg.fr/cgi-bin/cdsbib?2014yCat....1.2023S}, and was classified
as B0.5 in \cite{Rousseau} based upon low resolution objective prism plates.

\subsubsection{RX J0513.9-6951 Revisited}

\begin{figure}
\epsscale{1.0}
\plotone{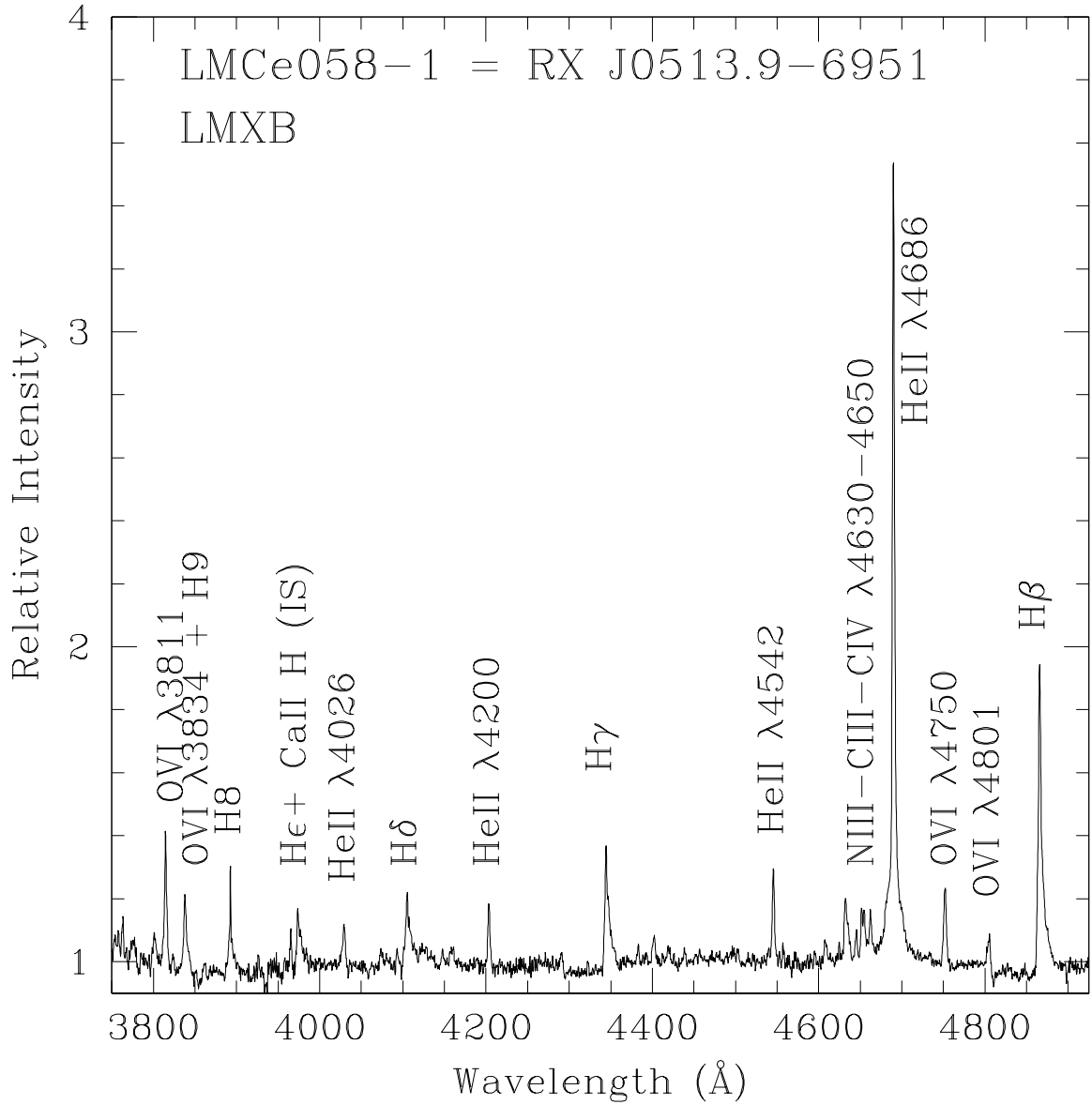}
\caption{\label{fig:LMCe0581} Spectrum of the LMXB LMCe058-1.  The principal emission lines are identified, and may be compared
to Fig.\ 2 in \cite{Cowley}. }
\end{figure}

One of the more interesting spectra we've encountered is that of LMCe058-1, shown in Fig.~\ref{fig:LMCe0581}.  The star is a known to be a super-soft x-ray
source,  RX J0513.9-6951, and its optical spectrum has been previously described by \cite{Cowley}.  The star is variable both in the x-ray region \citep{Cowley} as well as in the optical \citep{Leavitt,Cowley,MACHO}, with a Harvard Variable designation, HV 5682.   \cite{Cowley} describe the spectrum as typical of that of a low-mass X-ray binary (LMXB), with narrow emission components of He\,{\sc ii} and the Balmer lines, along with N\,{\sc iii} and C\,{\sc iii} and O\,{\sc vi} \citep[see also][]{Pakull}. The He\,{\sc ii} $\lambda$4686 profile clearly has two components, a narrow component and a broad component.  \cite{Cowley} attributes the broad component
to the high velocities of the inner part of the accretion disk.  \cite{Cowley} and \cite{Pakull} both note that the optical spectrum is very similar to that of Cal~83, the prototype of the LMXBs \citep{CramptonCal83}.

\begin{figure}
\epsscale{1.0}
\plotone{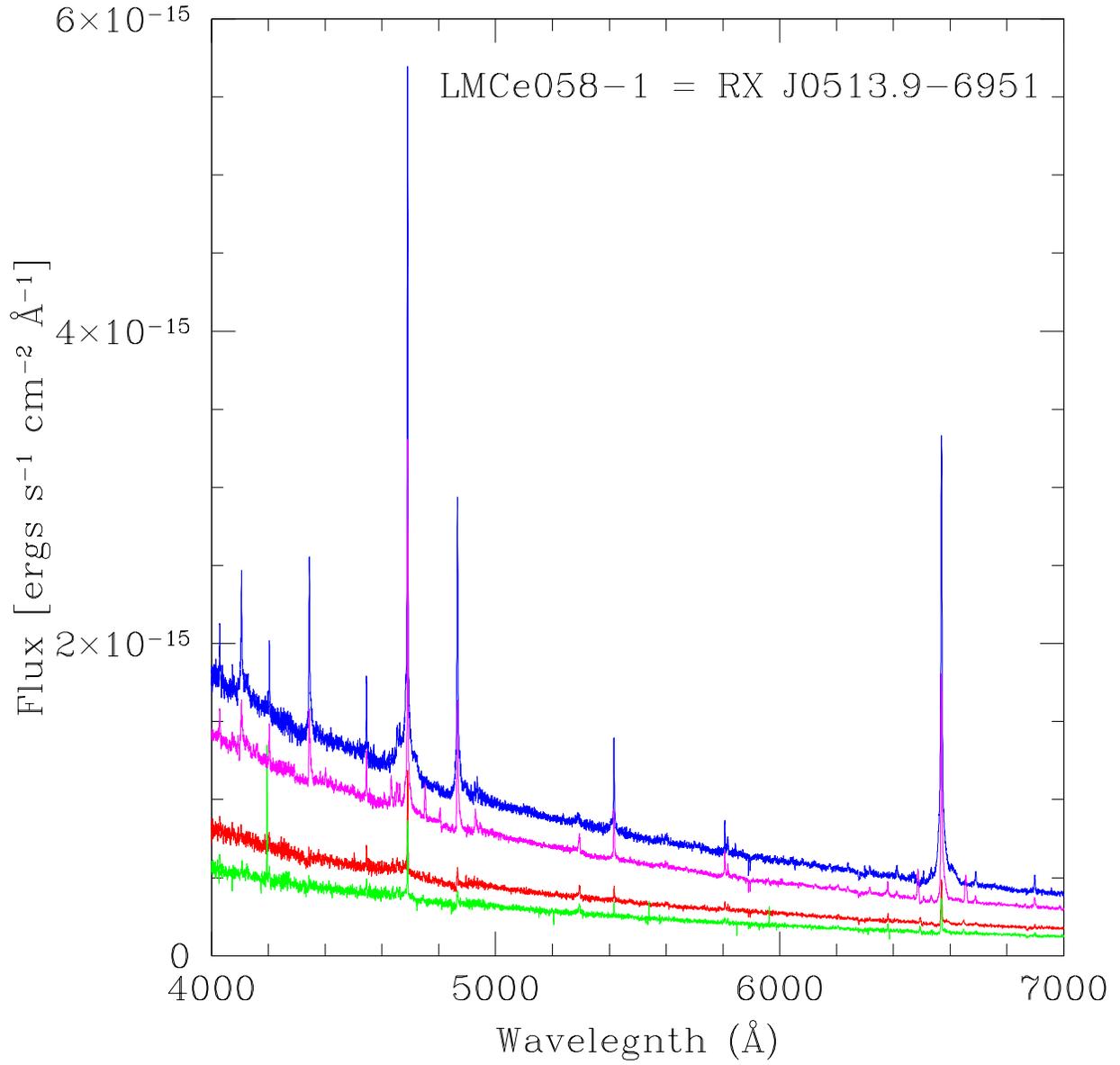}
\caption{\label{fig:LMCe058f} Spectral variability of the LMXB LMCe058-1.  From bottom to top, the spectra were taken on 2016 Feb 21, 2016 Feb 20, 2016 Jan 11, and 2016 Mar 29.}
\end{figure}

We reobserved this star as we were intrigued by the description of a broad He\,{\sc ii} $\lambda$4686 emission component.  Might this be a previously unrecognized Wolf-Rayet star with a disk?  Alas, the early descriptions are an accurate match to our optical spectrum; although He\ {\sc ii} $\lambda$4686 has a broad component, there are no other WR signatures present.  Qualitatively, the optical spectrum is very similar to that described over twenty years ago; compare Fig.~2 in \cite{Cowley} to our MagE spectrum in Fig.~\ref{fig:LMCe0581}.  We obtained four spectra of this star throughout a four month period and found dramatic changes in the emission line equivalent width; for instance, the EW of He\,{\sc ii} ranged from -5\AA\ to -25\AA. (\citealt{Cowley} reported an EW of -16\AA.)  Is this variability due to changes in the emission line fluxes or just to the change in the continuum level affecting the EWs?  To answer this, we turned to the fluxed
versions of our spectra.  As shown in Fig.~\ref{fig:LMCe058f}, it is clear that although the continuum flux varies, the emission line fluxes vary even more; the
two appear to be correlated, in that the emission lines are their strongest (both in terms of EWs and fluxes) when the star is brightest.

\subsection{Non-Emission-Line Stars}

In assigning observing priorities, we ranked our candidates on a 1-4 scale, with 1 being the most potentially interesting.
All of the emission-lined stars discussed above had been classified as a 1 or a 2.  All of our remaining candidates that were observed
were ranked as either a 3 or a 4,
and none of them proved to have He\,{\sc ii} $\lambda$4686 emission.  These priority 3-4 candidates were invariably B-type stars (including both a Be and B[e] star) plus one A0~I.  Given the large number of stars on our frames, and our desire to be thorough, a few low-significance candidates without emission are expected just on statistical grounds. Although these stars
are not of immediate interest to us, we list their properties in Table~\ref{tab:losers}.
These stars were classified using the relative strengths of Si\,{\sc iv} $\lambda$4089, Si\,{\sc iii} $\lambda$4553, Si\,{\sc ii} $\lambda$4128.  For stars later than B2, Mg\,{\sc ii} $\lambda$4481 and He\,{\sc i} $\lambda$4471 are secondary indicators.  (For more details, see \citealt{BigTable}.) In this, we made reference to the \cite{WF} atlas.  A difficulty, particularly for the SMC B stars, is that the metal lines are almost undetectable, and thus our classifications particularly uncertain even at our S/N. We confess to being influenced by our knowledge of the absolute visual magnitudes in assigning luminosity classes.

\section{Summary and Future Work}

We report here the detection and spectroscopic observations of 16 WR candidates from our late 2015/early 2016 observing season, the third year of our survey.  All six of the higher ranked candidates proved to have He\,{\sc ii} $\lambda$4686 emission.  Of these five, two are members of the WN3/O3 class, bringing to 10 the total number of this class.  All ten are in the LMC, and they make up $>$6\% of the LMC's WR known population, which now 
numbers 154.  The spectral features of nine of these WN3/O3 are remarkably homogeneous, but one of the newly found members here is not like the others, showing lower excitation emission spectrum (specifically, 
N\,{\sc iv} $\lambda$4058), P Cygni profiles for the N\,{\sc v} $\lambda\lambda$4503,19 emission, and weak He\,{\sc i} absorption.  Spectral modeling of all of these stars is in progress, with preliminary results reported by \cite{NeugentWN3O3}.

In addition, we have found three unusual Of-type stars, one of which is a member of the rare ``Ifc" luminosity class showing
relatively strong C\,{\sc iii} at $\lambda$4650, while two others are rapidly rotating ``Oef" type star (now often designated as
``nfp.")  We also present modern spectra of the low-mass x-ray binary RX J0513.9-6951, and note that qualitatively they look very similar to the discovery spectra of \cite{Cowley} and \cite{Pakull}.  Our repeated observations demonstrate that there are significant changes in the emission-line intensities, unsurprising given the star's large photometric variations. 

Ten of the candidates showed B-type or early A-type spectra; these were all lower ranked candidates and were observed for completeness.  One turned out to be a previously unrecognized Be star, and another a newly found B[e] star.

We are now prepared to begin the fourth year of our survey, with 26\% of the SMC and 12\% of the LMC left to observe.  We do not know what our final year of discoveries will bring us, but if past performance is any indication of future results, we are sure to find something interesting!

\acknowledgements
We thank the referee, Dr. Nolan Walborn, for useful comments which improved this paper.  We are grateful for the excellent support we always receive at Las Campanas Observatory, as well as the generosity of the Carnegie Observatory and Steward Observatory Arizona Time Allocation Committees.    Support for this project was provided by the National Science Foundation through AST-1008020 and AST-1612874, and through Lowell Observatory.   This research has made use of the VizieR catalogue access tool, CDS, Strasbourg, France. The original description of the VizieR service was published in A\&AS 143, 23.   The Catalog of Stellar Spectral Classification prepared by our colleague at Lowell Observatory, Brian Skiff, proved particularly useful, as always.  We also made use of data products from the Two Micron All Sky Survey (2MASS),  which is a joint project of the University of Massachusetts and the Infrared Processing and Analysis Center/California Institute of Technology, funded by the National Aeronautics and Space Administration and the NSF.   

{\it Facilities:} \facility{Magellan: Baade (MagE spectrograph)}, \facility{Swope (e2v imaging CCD)}

\clearpage

\bibliographystyle{apj}
\bibliography{masterbib}

\begin{deluxetable}{l c c c c c c r c c c l l l}
\rotate
\tabletypesize{\scriptsize}
\tablecaption{\label{tab:WRs} Newly Found Wolf-Rayet Stars}
\tablewidth{0pt}
\tablehead{
\colhead{ID\tablenotemark{a}}
& \colhead{$\alpha_{\rm 2000}$} 
& \colhead{$\delta_{\rm 2000}$} 
& \colhead{$V$\tablenotemark{b}}
&\colhead{$B-V$\tablenotemark{b}}
& \colhead{{\it CT}} 
& \multicolumn{2}{c}{\it WN - CT}
&
& \multicolumn{2}{c}{He\,{\sc ii} $\lambda$4686} 
& \colhead{$M_V$\tablenotemark{c}}
& \colhead{Sp.\ Type} 
& \colhead{Comment} \\  \cline{7-8} \cline{10-11}
&&&&&
&\colhead{mag} 
& \colhead{$\sigma$}
& \colhead{}  
& \colhead{$\log$(-EW\tablenotemark{d}} 
& \colhead{FWHM (\AA)} 
}
\startdata   
LMCe078-3 & 05 41 17.50 & $-69$ 06 56.2 & 17.03  & +0.03 & 17.2 & $-0.27$ & 12.0 && 1.2 & 16 & $-2.2$\tablenotemark{e} & WN3/O3 \\
LMCe055-1& 04 56 48.72 & $-69$ 36 40.3 & 16.15 & $-$0.10 & 16.4 & $-0.17$ & 8.4 && 0.8   & 17 & $-2.8$ &  WN4/O4 & OGLE LMC-ECL-3548 \\
\enddata
\tablenotetext{a}{Designation from the current survey.  We have denoted the e2v fields with a small ``e" to distinguish them from our numbering system from Paper I, i.e., LMCe159 is distinct from LMC159.  We plan to impose less idiosyncratic designations once our survey is complete.}
\tablenotetext{b}{Photometry from \citealt{DennisLMC}.}
\tablenotetext{c}{We assume an apparent distance modulus of 18.9 for the LMC, corresponding to a distance of 50 kpc \citep{vandenbergh} and an average extinction of $A_V=0.40$ \citep{MasseyLang,vandenbergh}.}
\tablenotetext{d}{``EW" is the equivalent width, measured in \AA.}
\tablenotetext{e}{Assumes an extra 0.3~mag of extinction at $V$ given its $B-V$ color.}
\end{deluxetable}

\begin{deluxetable}{l c c c c c c r c c c l l  l}
\rotate
\tabletypesize{\scriptsize}
\tablecaption{\label{tab:Ofs} Other Emission-Lined Stars}
\tablewidth{0pt}
\tablehead{
\colhead{ID\tablenotemark{a}}
& \colhead{$\alpha_{\rm 2000}$} 
& \colhead{$\delta_{\rm 2000}$} 
& \colhead{$V$\tablenotemark{b}}
&\colhead{$B-V$\tablenotemark{b}}
& \colhead{{\it CT}} 
& \multicolumn{2}{c}{\it WN - CT}
&
& \multicolumn{2}{c}{He\,{\sc ii} $\lambda$4686} 
& \colhead{$M_V$\tablenotemark{c}}
& \colhead{Sp.\ Type} 
& \colhead{Comment} \\  \cline{7-8} \cline{10-11}
&&&&&
&\colhead{mag} 
& \colhead{$\sigma$}
& \colhead{}  
& \colhead{$\log$(-EW\tablenotemark{d})} 
& \colhead{FWHM (\AA)} 
}
\startdata  
LMCe078-1 & 05 37 29.63   &	-69 14 52.0  & 13.49 & $-0.02$ & 13.8& $-0.07$ & 3.3 && 0.2 & 15 &  $-5.4$ & O6 Ifc & O5.5 Iaf in \cite{Evans} \\
LMCe078-2 & 05 37 27.84   &   -69 23 52.9  & 13.38 & $-0.16$ & 13.3& $-0.06$ & 3.2 && -0.2 & \nodata  & $-5.5$ & O5 nfp & \\
LMCe113-1 & 04 49 28.01   &   -67 42 39.8  & 13.18 & $-0.23$ & 13.2& $-0.09$ & 4.4 && -0.9 & \nodata  & $-5.7$ & O7.5nfp & \\
LMCe058-1 &  05 13 50.80 & $-69$ 51 47.6 & 16.71 & $-0.05$ &17.1& $-0.35$ & 15.3 &&  0.7\tablenotemark{e}  & 14\tablenotemark{d} & $-2.2$ & LMXB & RX J0513.9-6951 \\
\enddata
\tablenotetext{a}{Designation from the current survey.  We have denoted the e2v fields with a small ``e" to distinguish them from our numbering system from Paper I, i.e., LMCe159 is distinct from LMC159.  We plan to impose less idiosyncratic designations once our survey is complete.}
\tablenotetext{b}{Photometry from \citealt{DennisLMC}.}
\tablenotetext{c}{We assume an apparent distance modulus of 18.9 for the LMC, corresponding to a distance of 50 kpc \citep{vandenbergh} and an average extinction of $A_V=0.40$ \citep{MasseyLang,LGGSII}.}
\tablenotetext{d}{``EW" is the equivalent width, measured in \AA.}
\tablenotetext{e}{The He\,{\sc ii} $\lambda$4686 line consists of a broad component (WN star?) and a narrow component (disk?).  The values given
in the table are for the broad component.  The narrow component has a $\log$(-EW) of 0.8 and a FWHM of 2.8\AA.  The total $\log$(-EW) of the line is 1.0.}
\end{deluxetable}

\begin{deluxetable}{l c c c c c l  l }
\rotate
\tabletypesize{\scriptsize}
\tablecaption{\label{tab:losers} Stars without He\,{\sc ii} $\lambda$4686 Emission}
\tablewidth{0pt}
\tablehead{
\colhead{ID\tablenotemark{a}}
& \colhead{$\alpha_{\rm 2000}$} 
& \colhead{$\delta_{\rm 2000}$} 
& \colhead{$V$\tablenotemark{b}}
&\colhead{$B-V$\tablenotemark{b}}
& \colhead{$M_V$\tablenotemark{c}}
& \colhead{Sp.\ Type} 
& \colhead{Comment}
}
\startdata  
SMCe033-1 & 0:45:28.28	& -73:56:36.6 & 14.98 & $-0.14$ & $-4.1$&  B0.5~III & [M2002] SMC 5611;  B2 (II), \citealt{EvansSMC} \\
SMCe055-1 & 0:29:46.91 & -73:08:40.6 & 15.65 &  $-0.06$ & $-3.5$& B1-B1.5 IIIe & Weak Si~III, no Mg II.  Balmer emission \\
LMCe005-1 & 5:31:31.76 & -71:56:10.1 &  16.27 & $-0.07$ & $ -2.6$ & B2-V & Si~III and Mg~II\\
LMCe027-1 & 4:50:47.05 & -70:37:58.7 & 15.31 & $-0.10$ & $-3.6$ & B5~III &  [M2002] LMC 8808\\
LMCe029-1 & 4:59:01.99	& -70:48:53.0 & 15.37 & $-0.03$ & $-3.5$ & B0.5 V[e]& [M2002] LMC 41535; Balmer and Fe II (?) emission \\
LMCe050-2 & 5:46:05.71 & -69:59:57.2 & 16.65 & $+0.20$  & $-2.3$ & A0~V: & No He~I\\
LMCe117-1 & 5:06:13.53 & -67:49:11.3 & 15.96 & $+0.09$ & $-2.9$ & B2~V & Si~III and Mg~II\\
LMCe117-2 & 5:05:43.53	& -67:53:53.0 & 16.85 & $-0.05$ & $-2.0$ & B2~V & Si~III and Mg~II \\
LMCe141-1 & 4:51:19.42	& -66:52:18.9 & 16.90 & $+0.03$ & $-2.0$ & B8~III & Mg II stronger than He I  \\
LMCe155-1 & 5:03:37.35	& -66:33:19.5 & 16.42 & $-0.18$ & $-2.5$ & B2 V-III & Weak MgII and Si III\\
\enddata
\tablenotetext{a}{Designation from the current survey.  We have denoted the e2v fields with a small ``e" to distinguish them from our numbering system from Paper I, i.e., LMCe159 is distinct from LMC159.  We plan to impose less idiosyncratic designations once our survey is complete.}
\tablenotetext{b}{Photometry from \citealt{DennisSMC} for SMC members, and \citealt{DennisLMC} for LMC members.}
\tablenotetext{c}{For the SMC, we assume an apparent distance modulus of 19.1, corresponding to a distance of 59 kpc \citep{vandenbergh}, and an average extinction of $A_V=0.30$ \citep{MasseyLang,LGGSII}. For the LMC, we assume an apparent distance modulus of 18.9, corresponding to a distance of 50 kpc \citep{vandenbergh}, and an average extinction of $A_V=0.40$ \citep{MasseyLang,LGGSII}.}
\end{deluxetable}

\end{document}